\documentclass[journal]{IEEEtran}
\usepackage{amsmath, amssymb, amsthm}
\usepackage{accents}
\usepackage{bm}
\usepackage{commath}
\usepackage{graphicx}
\usepackage[caption=false,font=footnotesize]{subfig}
\usepackage[ruled,vlined,linesnumbered]{algorithm2e}
\usepackage[hidelinks]{hyperref}
\usepackage{doi}
\usepackage{xcolor}
\SetAlCapNameFnt{\small}
\SetAlCapFnt{\small}

\theoremstyle{definition}
\newtheorem{theorem}{Theorem}
\newtheoremstyle{named}{}{}{}{}{\bfseries}{.}{.5em}{#3 principle (#2)}
\theoremstyle{named}
\newtheorem{namedprinciple}{Principle}
\theoremstyle{definition}
\newtheorem{definition}{Definition}
\newtheorem{assumption}{Assumption}

\input{variables.sty}

\title{Structured learning of safety guarantees for the control of uncertain dynamical systems}

\author{Marc-Antoine Beaudoin,
	Benoit Boulet,~\IEEEmembership{Senior Member, IEEE}%
	\thanks{The authors are with the Intelligent Automation Lab of the Centre for Intelligent Machines and the Department of Electrical and Computer Engineering, McGill University, Montreal, QC, Canada, H3A 0E9 e-mail: (ma.beaudoin@mail.mcgill.ca, benoit.boulet@mcgill.ca).}%
	\thanks{Manuscript received December 06, 2021; revised January 21, 2022; accepted January 29, 2022. This work was supported by Mitacs and Quebec's Fonds de recherche Nature et technologies.}}%

\begin{document}
\maketitle

\begin{abstract}
	Approaches to keeping a dynamical system within state constraints typically rely on a model-based safety condition to limit the control signals. In the face of significant modeling uncertainty, the system can suffer from important performance penalties due to the safety condition becoming overly conservative. Machine learning can be employed to reduce the uncertainty around the system dynamics, and allow for higher performance. In this article, we propose the safe uncertainty-learning principle, and argue that the learning must be properly structured to preserve safety guarantees. For instance, robust safety conditions are necessary, and they must be initialized with conservative uncertainty bounds prior to learning. Also, the uncertainty bounds should only be tightened if the collected data sufficiently captures the future system behavior. To support the principle, two example problems are solved with control barrier functions: a lane-change controller for an autonomous vehicle, and an adaptive cruise controller. This work offers a way to evaluate whether machine learning preserves safety guarantees during the control of uncertain dynamical systems. It also highlights challenging aspects of learning for control.
\end{abstract}


\section{Introduction}
For a large class of dynamical systems---e.g., autonomous vehicles and unmanned aerial vehicles---the notion of safety implies keeping the system within state constraints. The general approach is to use a model of the system dynamics, and derive a condition on the control signals that guarantees system safety. Three such methods are: model predictive control (MPC)\cite{camacho_model_2007}, control barrier function (CBF)\cite{ames_control_2017}, and Hamilton-Jacobi reachability analysis (HJR)\cite{chen_hamiltonjacobi_2018}. 

In MPC, the control signal is optimized at every time step by predicting the future system states for a finite time-horizon. This optimization can account for state constraints, and therefore keep the system safe~\cite{falcone_predictive_2007,gao_predictive_2010,liniger_optimization-based_2015}. A CBF is a Lyapunov-like scalar function whose output depends on whether the system states are within a given set, and how far they are from the set boundary. A CBF and its time derivatives can be used to obtain a condition on the control signals for the system to remain within the given set, thereby enforcing forward invariance on that set. If the chosen set is within state constraints, then the CBF can be used to keep the system safe~\cite{nguyen_exponential_2016,ames_control_2019}. In HJR, a target set is defined as the set of states that violate the constraints, and a backward reachable (BR) set is defined as the set of states that could lead the system to the target set despite the system's best possible actions. The BR set, the optimal action, and a value function can all be computed by solving the Hamilton-Jacobi partial differential equation for this system, typically through dynamic programming. The system can be kept safe by applying the optimal action whenever the system comes close to the BR set~\cite{fisac_reach-avoid_2015,chen_reachability-based_2017}. These methods can also be combined: for instance, the value function obtained though HJR can be used as a CBF~\cite{choi_robust_2021}. \\

For systems with uncertain dynamics, all three methods outlined above can be adapted to consider bounded system uncertainty~\cite{bemporad_robust_1999,seo_robust_2019,nguyen_robust_2021}. Generally, this is nontrivial and often leads to a higher computational burden for the controller, since the controller must now predict the system evolution considering the set of possible system dynamics. Also, this can lead to overly conservative controllers if the uncertainty description is itself too conservative. For instance, this can arise if the method is formulated for a rather general uncertainty description, and the actual system uncertainty is in fact more restricted---more structured. Naturally, this motivates using machine learning to adapt the system model from data collected during the system's operation. An approach is where instead of using uncertainty bounds, a nominal model is learned from data, which reduces the risk associated with a discrepancy between the system model and the actual dynamics~\cite{taylor_learning_2020}. Another approach is considering uncertainty in the system dynamics and learning the uncertainty bounds from data, thereby increasing the performance of the system by obtaining tighter bounds~\cite{taylor_adaptive_2020,lopez_robust_2021,fisac_general_2019,hewing_cautious_2020,hewing_learning-based_2020}. \\

However, using machine learning to derive safety conditions can lead to unsafe scenarios if the model uncertainty is not learned properly. In other words, it is possible to follow the learning-based MPC, CBF, or HJR methods proposed in the literature and still render the system unsafe by not handling the uncertainty properly. As a complement to these methods, this article introduces the following principle.
\begin{namedprinciple}[Safe uncertainty-learning] \label{prop:1}
	Take an uncertain dynamical system with safety defined by state constraints, which must be respected at all time during operation. Suppose machine learning is used to obtain a condition on the control signal that will guarantee the system's safety. For the learning-based control method to preserve safety guarantees:
	\begin{enumerate}
		\item A robust safety condition is necessary, where uncertainty bounds are considered.
		\item The uncertainty bounds must be initialized conservatively. 
		\item The uncertainty bounds should only be tightened if it can be assumed that the collected data sufficiently captures the future behavior of the system. Particularly challenging adversarial events are a sudden change in the system dynamics or a rare disturbance. If expected, such events must be distinctly accounted for, since it could be impossible to model them from previously collected data.  
	\end{enumerate}
\end{namedprinciple}
Our argument is based on contraposition: through two examples, we show that when one of these conditions is not respected, the safety guarantee is lost. The examples are 1) the lateral control of an autonomous vehicle through a lane-change maneuver, and 2) the longitudinal control of an autonomous vehicle in a two-vehicle platoon, also commonly called adaptive cruise control. Both these problems were investigated numerous times in the literature, including with some of the approaches presented above~\cite{ames_control_2017, gao_predictive_2010, gattami_establishing_2011}. In this work, we use robust exponential control barrier functions. Also, we learn parametric uncertainty with maximum likelihood estimation. Section~\ref{sec:approach} presents the chosen approaches to obtain the safety condition and to learn the parameters. Sections~\ref{sec:example1}~and~\ref{sec:example2} present both example problems and the simulation results. Finally, Section~\ref{sec:discussion} discusses Principle~\ref{prop:1} in light of the example results, as well as other approaches from the literature. Notably, the implications of non-parametric models are discussed.  

\section{Learned safety condition} \label{sec:approach}
This section presents the approach to derive the learned safety condition for the example problems of Sections~\ref{sec:example1}~and~\ref{sec:example2}. It begins with the introduction of CBFs, then exponential-CBFs. Uncertainty is then considered and robust exponential-CBFs are introduced. The definitions and theorems are largely adapted from~\cite{ames_control_2019,lopez_robust_2021}. Finally, the chosen learning method is presented---i.e., maximum likelihood estimation of the uncertainty parameters. 
\subsection{Control barrier function} \label{sec:cbf}
Take a system with states $\bx \in \mathbb{R}^n$, controls $\bu \in \mathbb{R}^m$, and dynamics 
\begin{equation}
	\bxp = f(\bx) + g(\bx)\bu. \label{eq:dynamics}
\end{equation}
If the dynamics are locally Lipschitz, then given an initial condition $\bx_0$, there exists a maximum time interval ${I(\bx_0) = [t_0,T)}$ such that $\bx(t)$ is a unique solution on $I(\bx_0)$~\cite{ames_control_2019}. Let a closed convex set $\calS \subset \mathbb{R}^n$ defined as the 0-superlevel set of a continuously differentiable function $h:\mathbb{R}^n \rightarrow \mathbb{R}$ where   
\begin{align}
	\calS &:= \{ \bx \in \mathbb{R}^n | h(\bx) \geq 0 \}, \\
	\partial \calS &:= \{ \bx \in \mathbb{R}^n | h(\bx) = 0 \}, \\
	\interior(\calS) &:= \{ \bx \in \mathbb{R}^n | h(\bx) > 0 \}.
\end{align}
\begin{definition}
	The set $\calS$ is forward invariant if for every ${\bx_0 \in \calS}$, $\bx(t) \in \calS \ \forall  t \in I(\bx_0)$.
\end{definition}
\begin{definition}
	A continuous function $\alpha: \mathbb{R} \rightarrow \mathbb{R}$ is an extended class $\Kinf$ function if it is strictly increasing, $\alpha(0) = 0$, and is defined on the entire real line.
\end{definition}
\begin{definition}
	Let $\calS$ be a 0-superlevel set for a continuously differentiable function $h:\mathbb{R}^n \rightarrow \mathbb{R}$, then $h$ is a control barrier function if there exists an extended class $\Kinf$ function $\alpha$ such that for all $\bx \in \calS$
	\begin{equation}
		\sup_{\bu\in\mathcal{U}} \dot{h}(\bx,\bu) \geq -\alpha(h(\bx)). \label{eq:h}
	\end{equation} \label{def:cbf}
\end{definition}
\begin{theorem} \label{thm:safe}
	Let $\calS$ be a 0-superlevel set for a continuously differentiable function $h:\mathbb{R}^n \rightarrow \mathbb{R}$, if $h$ is a control barrier function for \eqref{eq:dynamics} on $\calS$, then any Lipschitz continuous controller satisfying \eqref{eq:h} renders the set $\calS$ forward invariant~\cite{ames_control_2019}.
\end{theorem}
As a result of Theorem~\ref{thm:safe}, if the set $\calS$ is defined as the safe set, then any controller that meets Condition~\eqref{eq:h} keeps the system safe. 

\subsection{Exponential control barrier function} \label{sec:ecbf}
For systems where $\dot{h}(\bx)$ does not depend on $\bu$, Definition~\ref{def:cbf} cannot be used to find a control input $\bu$ that keeps the system safe. It is possible, however, to use higher derivatives of $h(\bx)$. For a system where the $r$\textsuperscript{th} time derivative of $h(\bx)$ depends on $\bu$, but not any of the lower derivatives, a new system with states $\eta(\bx):= [h(\bx),\dot{h}(\bx),\ddot{h}(\bx),\ldots,h^{(r-1)}(\bx)]^\top$, input ${\mu = h^{(r)}(\bx,\bu)}$, and output $h(\bx)$ can be formed. The dynamics of this system are 
\begin{align}
	\dot{\eta}(\bx) &= F\eta(\bx) + G\mu, \label{eq:etadot}\\
	h(\bx) &= C\eta(\bx),
\end{align}
with 
\begin{align}
	F &= \begin{bmatrix}
		0 & 1 & 0 & \cdots & 0 \\
		0 & 0 & 1 & \cdots & 0 \\
		\vdots & \vdots & \vdots & \ddots & \vdots \\
		0 & 0 & 0 & \cdots & 1 \\
		0 & 0 & 0 & \cdots & 0
	\end{bmatrix}, \quad
	G = \begin{bmatrix}
		0 \\ 0 \\ \vdots \\ 0 \\ 1
	\end{bmatrix}, \nonumber \\
	C &= \begin{bmatrix}
		1 & 0 & 0 & \cdots & 0
	\end{bmatrix}. \label{eq:etamat}
\end{align}
If $\mu = -K_\alpha \eta(\bx)$ is chosen, ${h(\bx(t)) = C e^{(F-GK_\alpha)t}\eta(\bx_0)}$. By the comparison lemma~\cite{ames_control_2019}, if $\mu \geq -K_\alpha \eta(\bx)$, then ${h(\bx(t)) \geq C e^{(F-GK_\alpha)t}\eta(\bx_0)}$. 
\begin{assumption}
	$K_\alpha$ is chosen such that the poles $p_i, \ {i\in\{1,\ldots,r\}}$ of $(F-GK_\alpha)$ are real and negative. \label{ass:Ka1}
\end{assumption}
\begin{assumption}
	Define a family of recursive functions ${\nu_i: \mathbb{R}^n \rightarrow \mathbb{R}}, \ i\in\{0,\ldots,r\}$, and corresponding superlevel sets $\calS_i$ as 
	\begin{align}
		\nu_i(\bx) &= \dot{\nu}_{i-1}(\bx) - p_i \nu_{i-1}(\bx),\\
		\calS_i &= \{ \bx \in \mathbb{R}^n \ |\ \nu_i(\bx) \geq 0\},
	\end{align}
	and define $\nu_0(\bx) := h(\bx)$. Then, $K_\alpha$ is chosen such that $\nu_i(\bx_0) \geq 0, \ i\in\{1,\ldots,r\}$. \label{ass:Ka2}
\end{assumption} \newpage

\begin{definition}
	Let $\calS$ be a 0-superlevel set for a $r$-times continuously differentiable function ${h:\mathbb{R}^n \rightarrow \mathbb{R}}$, then $h$ is an exponential control barrier function if there exists a row vector $K_\alpha \in \mathbb{R}^r$ constrained by  Assumptions~\ref{ass:Ka1}~and~\ref{ass:Ka2}, such that for all $\bx \in \calS$
	\begin{equation}
		\sup_{\bu\in\mathcal{U}} h^{(r)}(\bx,\bu)  \geq -K_\alpha \eta(\bx). \label{eq:hr}
	\end{equation}  \label{def:ecbf}
\end{definition}

\subsection{Robust exponential control barrier function} \label{sec:recbf}
For systems where the dynamics are uncertain, Definition~\ref{def:ecbf} is inappropriate to find a control input $\bu$ that keeps the system safe. Take an uncertain system model where the dynamics depend explicitly on an unknown parameter vector $\bd$, with each parameter bounded in magnitude such that $\bd \in \mathcal{D}$, where
\begin{equation}
	\bxp = f(\bx,\bd) + g(\bx, \bd)\bu.
\end{equation}
Now, the control barrier function $h(\bx,\bd)$ and its derivatives depend on the unknown parameters~$\bd$. This requires to adapt Assumption~\ref{ass:Ka2} and Definition~\ref{def:ecbf}.
\begin{assumption}
	Define a family of recursive functions ${\nu_i: \mathbb{R}^n \rightarrow \mathbb{R}}, \ i\in\{0,\ldots,r\}$, and corresponding superlevel sets $\calS_i$ as 
	\begin{align}
		\nu_i(\bx,\bd) &= \dot{\nu}_{i-1}(\bx,\bd) - p_i \nu_{i-1}(\bx,\bd), \\
		\calS_i &= \{ \bx \in \mathbb{R}^n \ |\ \nu_i(\bx,\bd) \geq 0\},
	\end{align}
	and define $\nu_0(\bx,\bd) := h(\bx,\bd)$. Then, $K_\alpha$ is chosen such that $\nu_i(\bx_0,\bd) \geq 0, \ i\in\{1,\ldots,r\}, \ \forall \bd \in \mathcal{D}$. \label{ass:Ka3}
\end{assumption}
\begin{definition}
	Let $\calS$ be a 0-superlevel set for a $r$-times continuously differentiable function ${h:\mathbb{R}^n \rightarrow \mathbb{R}}$, then $h$ is a robust exponential control barrier function if there exists a row vector $K_\alpha \in \mathbb{R}^r$ constrained by  Assumptions~\ref{ass:Ka1}~and~\ref{ass:Ka3} such that for all $\bx \in \calS$ and for all $\bd \in \mathcal{D}$
	\begin{equation}
		\sup_{\bu\in\mathcal{U}} h^{(r)}(\bx,\bu,\bd)  \geq -K_\alpha \eta(\bx,\bd). \label{eq:rhr}
	\end{equation}  \label{def:recbf}
\end{definition}

\subsection{Robustly safe controller}
The robustly safe controller used in this work consists of correcting a nominal $\kn(\bx)$ controller's output whenever Condition~\eqref{eq:rhr} is not respected. This can be expressed as the following optimization problem.
\begin{align}
	\bu = &\argmin_{\bu\in\mathcal{U}} \ \tfrac{1}{2}\Vert \bu-\kn(\bx) \Vert_2^2, \\
	\mathrm{s.t.} \ &h^{(r)}(\bx,\bu,\bd)  \geq -K_\alpha \eta(\bx,\bd), \ \forall \bd \in \mathcal{D}.
\end{align}

This optimization problem is in general nontrivial. It may be possible to further structure the problem depending on the system equations, the uncertainty description, and the chosen control barrier function.

\subsection{Learning parametric uncertainty} \label{sec:sysID}
When the controller is initialized, conservative bounds $\mathcal{D}$ are estimated for the parameters $\bd$. As the controller is deployed and data is collected, it may be possible to update the bounds for $\bd$. In this work, the following structure is assumed
\begin{align}
	\bxp &= f(\bx,\btheta) + g(\bx, \btheta)\bu, \\
	\by &= l(\bx,\btheta), \\
	\bz_{[n]} &= \by_{[n]} + G \bw_{[n]}, \\
	\bw_{[n]} &\sim \mathcal{N}\big(\bm{0},\Sigma_{\bw}\big),
\end{align}
where $\btheta$ are uncertain model parameters, which include the parameters $\bd$ of interest, $\by$ are the system output, $\bz_{[n]}$ are the $n$-indexed measurements contaminated by Gaussian noise $\bw_{[n]}$. The noise is assumed to be zero-mean and to have a known and diagonal covariance matrix $\Sigma_{\bw}$. To obtain the best estimate for the parameters $\btheta$, the likelihood of the observed data $p(z|\btheta)$ is maximized. For that, $p(z|\btheta)$ is assumed Gaussian. Maximizing $p(z|\btheta)$ is then equivalent to maximizing 
\begin{equation}
	J(\btheta) = \frac{1}{2}\sum_{n=1}^{N} (\bz_{[n]}-\by_{[n]})^\top \Sigma_{\bw}^{-1} (\bz_{[n]}-\by_{[n]}) ,
\end{equation}
where $N$ is the number of data point collected. In this work, $J(\btheta)$ is maximized through gradient ascent. Finally, the following covariance matrix is used to estimate the uncertainty bounds around the parameters
\begin{equation}
	P = \left\{\sum_{n=1}^{N} \left[ \dod{\by_{[n]}}{\btheta} \right]^\top \Sigma_{\bw}^{-1} \left[ \dod{\by_{[n]}}{\btheta} \right] \right\}^{-1}.
\end{equation}

This method is commonly used in practice to identify system parameters, such as flight dynamics~\cite{jategaonkar_flight_2015}. A notable caveat, however, is that the recorded data is not guaranteed to provide enough information to identify the parameters with enough accuracy to be useful for the given application. This is mitigated by the fact that we also estimate uncertainty bounds around the parameters, and can therefore decide not to update the parameters when the identification is not precise enough, i.e., when the variance of the parameter estimation in $P$ is too high. 

\section{Example 1: vehicle lateral control} \label{sec:example1}
The first example problem is to control a lane change maneuver performed by an autonomous vehicle. The goal is to bring the vehicle form the middle of a lane where the lateral position is defined as $Y$ = 0\,m, to the middle of a neighboring lane where $Y$ = 3.7\,m. However, the vehicle must not exceed $\Ymax$ = 3.85\,m in order to avoid a potential collision with other vehicles. In this problem, the vehicle speed is assumed to remain constant throughout the maneuver. 

\subsection{System model \& uncertainty}
\begin{figure}
	\centering
	\includegraphics[width=0.32\linewidth,trim={2.4cm 19.0cm 14.5cm 1.6cm},clip]{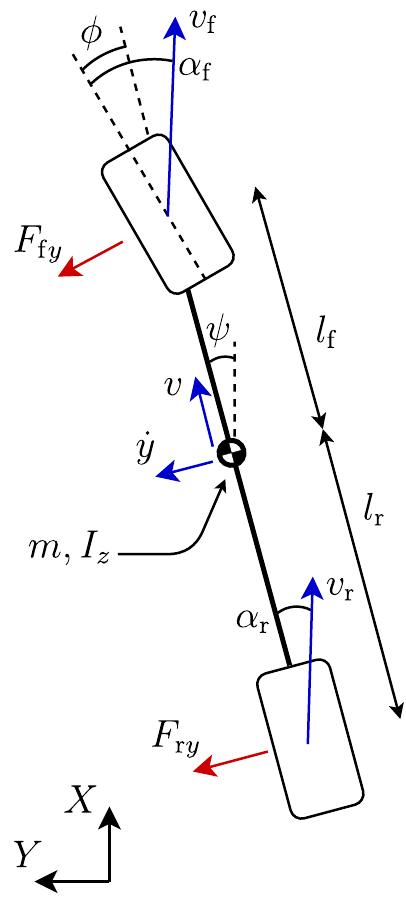}
	\caption{Bicycle model used in Example 1.}
	\label{fig:fbd}
\end{figure}

The vehicle is modeled with a bicycle model as shown on Figure~\ref{fig:fbd}. The states of the system are $\bx = [\yp,\psip, \psi, Y, \phi]^\top$, and the control is $\bu = [\phir]$. The dynamics are as shown below; full state measurement is assumed. 
\begin{align}
	\bxp &= A \bx + B \bu. \label{eq:dyna} \\
	A &= \begin{bmatrix}
		\frac{-(\Cf + \Cr)}{m v_0} & \frac{-\Cf \lf + \Cr \lr}{m v_0} - v_0 & 0 & 0 & \frac{\Cf}{m}  \\
		\frac{-\Cf \lf + \Cr \lr}{I_z v_0} \rule{0pt}{3.5ex} & \frac{-(\Cf \lf^2 + \Cr \lr^2)}{I_z v_0} & 0 & 0 & \frac{\Cf \lf}{I_z} \\
		0 \rule{0pt}{2.5ex} & 1 & 0 & 0 & 0 \\
		1 & 0 & v_0 & 0 & 0 \\
		0 & 0 & 0 & 0 & -\lambdas
	\end{bmatrix}, \\
	B &= \begin{bmatrix}
		0 & 0 & 0 &	0 &	\lambdas
	\end{bmatrix}^\top. \label{eq:dyna_B}
\end{align}

The actual vehicle parameters are unknown to the controller, so parametric uncertainty is used to bound the system behavior, see Table~\ref{tbl:params2}. The vehicle is assumed to be a medium-duty truck, so its mass $m$ may vary significantly depending on the payload it carries. The vehicle's moment of inertia $\Iz$ is assumed to vary proportionally to the mass of the vehicle, as well as an unknown parameter $\dI$ that accounts for an unknown mass distribution. The wheelbase of the vehicle is known to be $l$ = 4.5\,m. The distance of the front axle to the center of mass is unknown, but assumed to vary around 55\% of the wheelbase. The front and rear tire cornering stiffness---$\Cf$ and $\Cr$, respectively---are assumed to be proportional to the weight on the given axle, as well as a parameter $c_\alpha$. This $c_\alpha$ parameter is dependent on the tire construction and its loading condition~\cite{ervin_effects_1976,tapia_extended_1983}, thus is parameterized as well. Finally, the steering mechanism is modeled as a first order system with a unknown decay constant $\lambdas$~\cite{you_vehicle_1998}.

This parametrization allows to express the $A$ and $B$ matrices in Equation~\eqref{eq:dyna} explicitly in terms of the uncertain parameters. Below are the non-zero matrix components, where $a_{ij}$ represents $[A]_{ij}$. As a result of the parametrization, $\dm$ does not appear in the terms.
\begin{align}
	&a_{11} = -\cn \delta_2g/v_0, \\
	&a_{12} = -v_0,  \\
	&a_{15} = \cn \delta_2 g (1-\an \delta_1), \\
	&a_{21} = 0,  \\
	&a_{22} = -\cn \delta_2 g l (1-\an \delta_1)\an \delta_1/(\dn \dI v_0), \\
	&a_{25} = \cn \delta_2 g (1-\an \delta_1)\an \delta_1/(\dn \dI),  \\
	&a_{55} = -\lambdan\delta_3, \\
	&b_{51} = \lambdan\delta_3. 
\end{align}

\begin{table}
	\centering
	\caption{Uncertainty parametrization for Example 1}\label{tbl:params2}
	\begin{tabular}{c l l | c c c }
		\multicolumn{2}{c}{Parametrization} & Unit     & Param.     & Nom. & Range     \\ \hline
		$m$     & $\mn \dm$              & kg       & $\mn$      & 6500 & -         \\
		$I_z$    & $\dn m \dI$            & kg\,m\sq & $\dn$      & 4.8  & -         \\
		$\lf$    & $\an l \delta_1$       & m        & $l$        & 4.5  & -         \\
		$\lr$    & $l-\lf$                & m        & $\an$      & 0.55 & -         \\
		$\Calpha$  & $\cn \delta_2$         & rad\mo   & $\cn$      & 8    & -         \\
		$\Cf$    & $\Calpha m g \lr/l$    & N/rad    & $\lambdan$ & 8    & -         \\
		$\Cr$    & $\Calpha m g \lf/l$    & N/rad    & $\dm$      & 1    & $\pm 0.3$ \\
		$\lambdas$ & $\lambdan \delta_3$    & s\mo     & $\dI$      & 1    & $\pm 0.3$ \\
		&                        &          & $\delta_1$ & 1    & $\pm 0.4$ \\
		&                        &          & $\delta_2$ & 1    & $\pm 0.4$ \\
		&                        &          & $\delta_3$ & 1    & $\pm 0.4$ \\ \hline
	\end{tabular}
\end{table}

\subsection{Safety controller}
The safety condition is turned into an exponential control barrier function $h(\bx)$, which has the following derivatives
\begin{align}
	h(\bx) &= \Ymax-Y, \\
	\hp(\bx) &= -\yp -v_0 \psi, \\
	\hpp(\bx) &= -a_{11} \yp - a_{15}\phi, \\
	\hppp(\bx,\bu) &= -a_{11}^2 \yp + a_{11}v_0 \psip \nonumber \\
	& \quad \quad- a_{15}(a_{11}+a_{55})\phi -a_{15}b_{51} \phir.
\end{align}

A system with states $\eta(\bx) = [h(\bx),\hp(\bx),\hpp(\bx)]^\top$, input $\mu = \hppp(\bx,\bu)$, and dynamics that follow Equations~\eqref{eq:etadot}--\eqref{eq:etamat} is composed. A controller $K_\alpha = [k_1,k_2,k_3]$ is devised with pole placement. As a result of Definition~\ref{def:recbf}, the safety of the system is ensured with the following condition
\begin{align}
	\phir &\leq \min_{\bd \in \mathcal{D}} s(\bx,\bd), \label{eq:LCcond} \\
	s(\bx,\bd) &:= \frac{1}{a_{15}b_{51}} \Big( k_1 \Ymax - \big( k_2 + a_{11} k_3 + a_{11}^2 \big) \yp \nonumber \\
	& \quad \quad + a_{11}v_0 \psip - k_2 v_0 \psi - k_1 Y \nonumber \\
	& \quad \quad \quad   -a_{15}(k_3 + a_{11} + a_{55}) \phi \Big), \\
	\bd &= [\delta_1, \delta_2, \delta_3]^\top.
\end{align} 
This safety condition is formulated as a direct limitation on the controller's output signal $\phir$. Therefore, a safe controller must ensure that Condition~\eqref{eq:LCcond} is respected at every time step. This optimization problem is not trivial, as the parameters in $\bd$ appear multiple times in $s(\bx,\bd)$. In this work, it was found sufficient to discretize the search space $\mathcal{D}$ and evaluate $s(\bx,\bd)$ at every point of the resulting 3D grid. With 10 points per axis, this requires 1000 function evaluations every time step, which is easily manageable in real-time. \\

The base controller is a linear state feedback controller.
\begin{equation}
	\kn(\bx) = K(\bxr - \bx),
\end{equation}
where $\bxr$ is the desired reference state, namely $\bxr = [0,0,0,3.7,0]^\top$. In this work, $K$ is obtained by solving a LQR problem. Finally, $\phir$ is saturated at 0.08\,rad to counter the large input resulting when the desired state is suddenly switched to 3.7\,m at $t$ = 0\,s. 

\subsection{Parameter identification} \label{sec:lblc}
In order to reduce the uncertainty bounds around the components of $\bd$, parameter identification is performed according to the method presented in Section~\ref{sec:sysID}, where
\begin{align}
	\by_{[n]} &= \bxp_{[n]}(\bx,\bu,\btheta), \\
	\bz_{[n]} &= \by_{[n]} + \bw_{[n]}, \label{eq:measLC} \\
	\Sigma_{\bw} &= 0.1^2 I_5, \label{eq:sigmaLC} \\
	\btheta &= [\dI,\delta_1,\delta_2,\delta_3]^\top.
\end{align}
Because the measurements are obtained from a simulated environment, random noise is injected in the measurements as per Equations~\eqref{eq:measLC}~and~\eqref{eq:sigmaLC} to have a more realistic identification problem. 

\subsection{Results \& discussion}
\begin{figure}
	\centering
	\subfloat[Nominal model.\label{fig:Yplot_a}]{\includegraphics[width=0.4\linewidth,trim={0.0cm 0.0cm 0.0cm 0.0cm},clip]{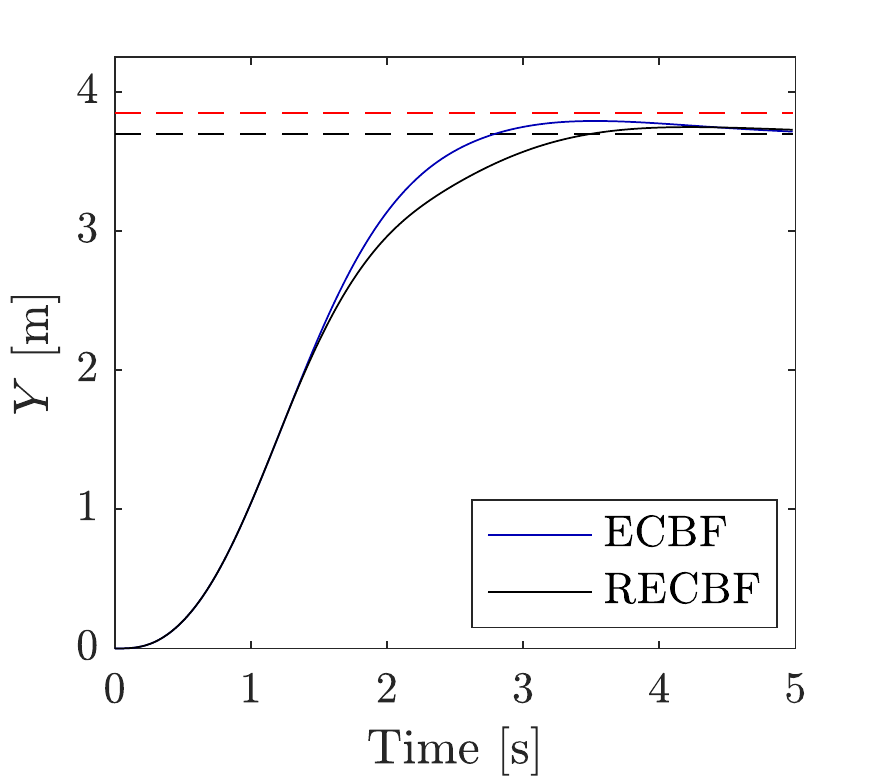}} \\
	\subfloat[Vehicle \#1.\label{fig:Yplot_1}]{\includegraphics[width=0.4\linewidth,trim={0.0cm 0.0cm 0.0cm 0.0cm},clip]{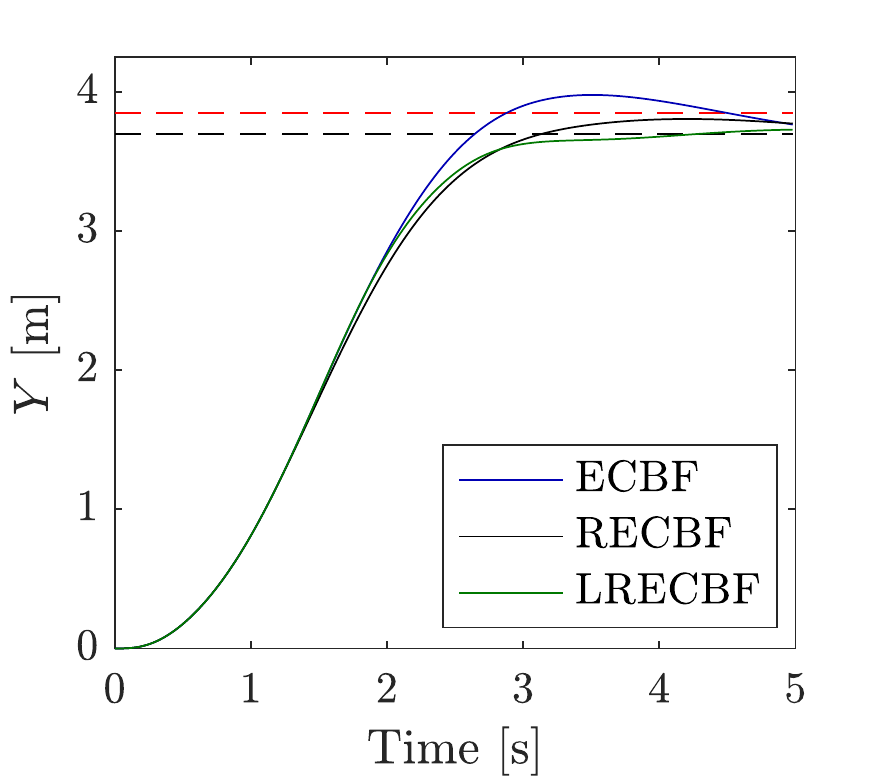}} \quad
	\subfloat[Vehicle \#2.\label{fig:Yplot_2}]{\includegraphics[width=0.4\linewidth,trim={0.0cm 0.0cm 0.0cm 0.0cm},clip]{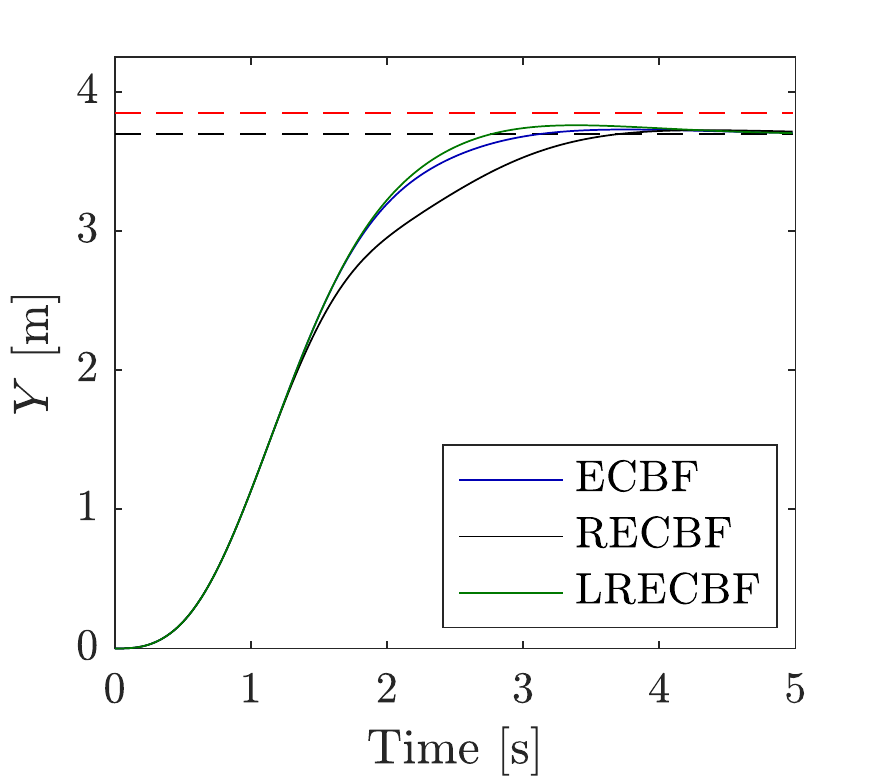}}
	
	\caption{Simulation results for the lane change maneuvers of Example 1.}
	\label{fig:Yplots}
\end{figure}

\begin{figure}
	\centering
	\subfloat[Control output.\label{fig:phir}]{\includegraphics[width=0.4\linewidth,trim={0.0cm 0.0cm 0.0cm 0.0cm},clip]{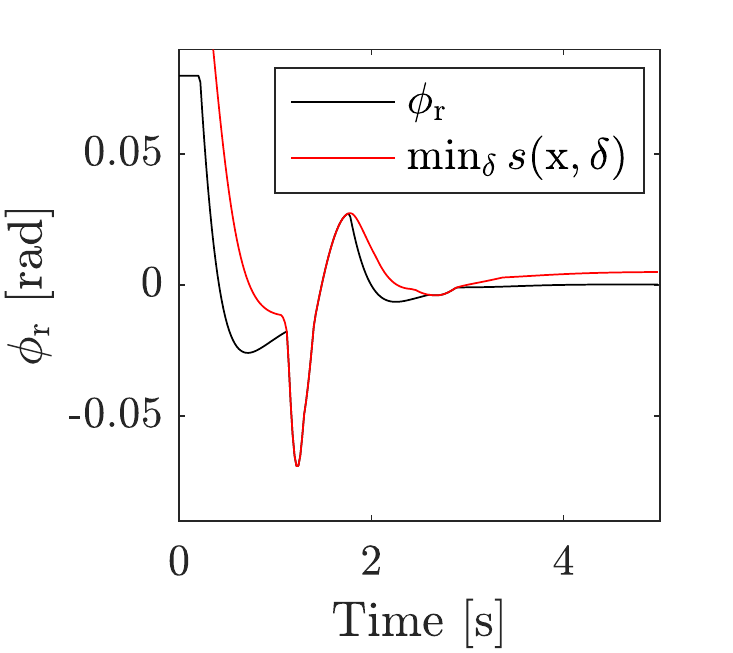}} \quad
	\subfloat[$\delta_1$ that minimizes $s(\bx,\bu)$.\label{fig:delta1}]{\includegraphics[width=0.4\linewidth,trim={0.0cm 0.0cm 0.0cm 0.0cm},clip]{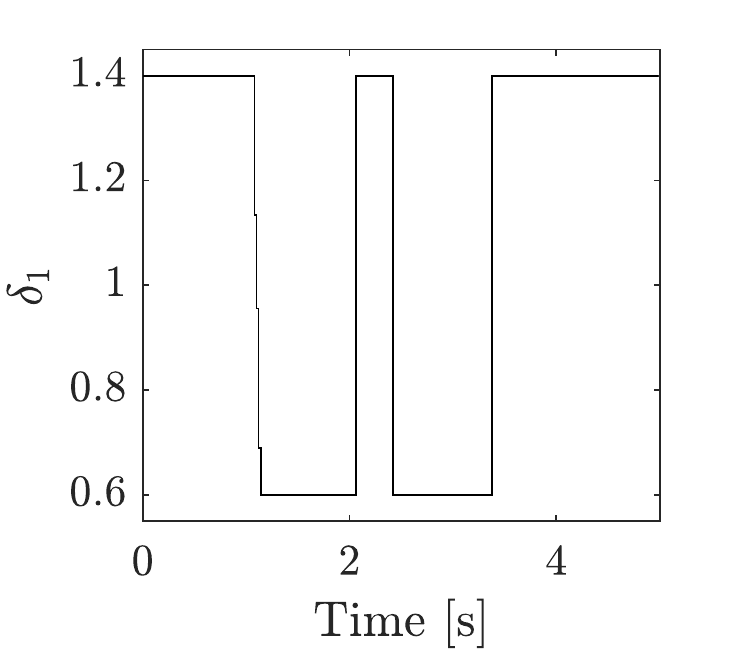}} \\
	\subfloat[$\delta_2$ that minimizes $s(\bx,\bu)$.\label{fig:delta2}]{\includegraphics[width=0.4\linewidth,trim={0.0cm 0.0cm 0.0cm 0.0cm},clip]{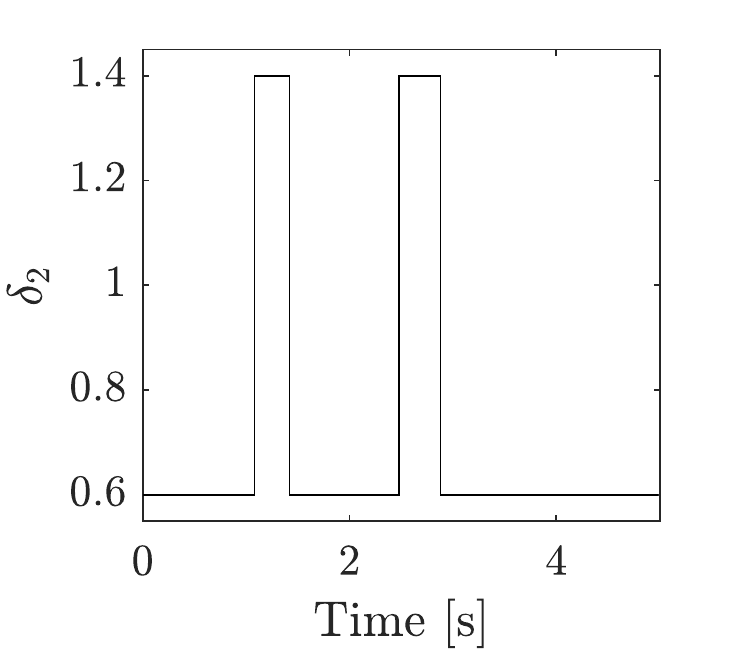}} \quad
	\subfloat[$\delta_3$ that minimizes $s(\bx,\bu)$.\label{fig:delta3}]{\includegraphics[width=0.4\linewidth,trim={0.0cm 0.0cm 0.0cm 0.0cm},clip]{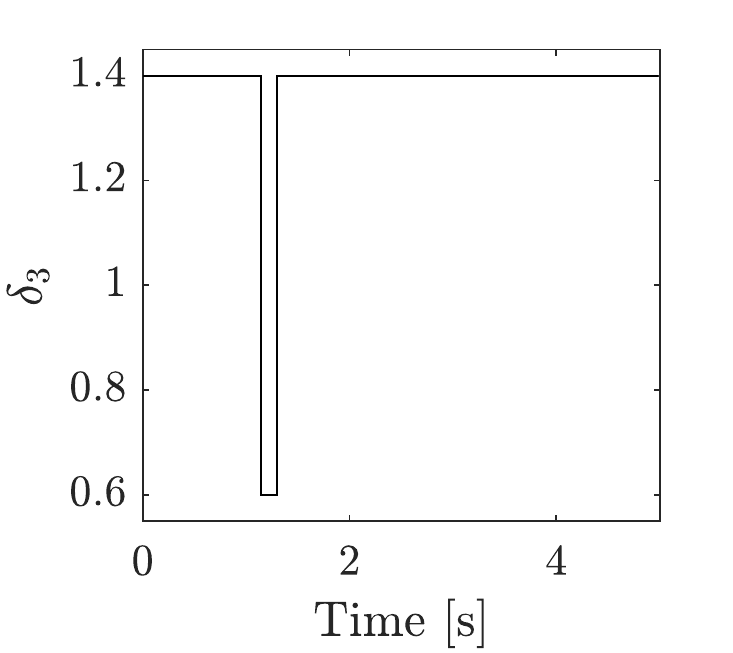}}
	
	\caption{Details of the \textsc{recbf} simulation for the nominal vehicle model in Figure~\ref{fig:Yplot_a}.}
	\label{fig:deltas}
\end{figure}

Figure~\ref{fig:Yplots} shows the lateral vehicle position of various simulated runs. The simulated vehicles are based on the system model of Equations~\eqref{eq:dyna}--\eqref{eq:dyna_B}. In Figure~\ref{fig:Yplot_a} the vehicle parameters are at nominal value, and in Figures~\ref{fig:Yplot_1}~and~\ref{fig:Yplot_2}, the vehicles have the parameters of Vehicles~\#1~and~\#2 respectively, which are listed in Table~\ref{tbl:pdeltas}. In all three plots, the blue lines originate from a safety controller that does not account for uncertainty bounds around the system parameters; it is an exponential control barrier function (\textsc{ecbf}) as presented in Section~\ref{sec:ecbf}. This is sufficient to keep the nominal vehicle safe, but not Vehicle~\#1. When considering uncertainty bounds around the system parameters and using the robust-\textsc{ecbf} presented in Section~\ref{sec:recbf}, all three systems remain safe. Figure~\ref{fig:phir} shows the effect of the \textsc{recbf} on the control action during the simulation with the nominal vehicle model. The black line represents the controller signal $\phir$, and the red line, the maximal value that $\phir$ could take and still respect Condition~\eqref{eq:LCcond}. When the two lines coincide, it means that the \textsc{recbf} limits the control action; otherwise, only the nominal controller $\kn(\bx)$ is in effect. Figures~\ref{fig:delta1}-\ref{fig:delta3} show the values of $\bd \in \mathcal{D}$ that minimize $s(\bx, \bd)$ at every time step of the \textsc{recbf} lane change in Figure~\ref{fig:Yplot_a}. These values are the solution of the minimization problem in Condition~\eqref{eq:LCcond}. Since the system states evolve with time and $s(\bx, \bd)$ is state-dependent, the minimization problem is repeated every time step. The plots show the worst possible perturbations of the system parameters $\bd \in \mathcal{D}$ according to $s(\bx, \bd)$. The values range from 0.6 to 1.4 as $\mathcal{D}$ is not yet learned and is therefore still defined by the initial uncertainty bounds of Table~\ref{tbl:params2}. Most of the time, these values are located at the boundary of $\mathcal{D}$, which supports the assumption that the rather crude optimization method chosen for this problem is adequate.

Figures~\ref{fig:Yplot_a}~and~\ref{fig:Yplot_2} show that for some systems, the robust formulation of the safety controller may hinder the system performance by delaying the completion of the lane change. This motivates learning the actual system parameters, and hopefully estimating smaller uncertainty bounds. The system parameters are estimated for the Vehicles~\#1~and~\#2 by collecting data issued from the \textsc{recbf} runs and then applying the method proposed in Section~\ref{sec:lblc}. The estimation results are shown in Table~\ref{tbl:pdeltas}, and the resulting lane change maneuvers are shown in green (\textsc{lrecbf}) in Figure~\ref{fig:Yplot_1}~and~\ref{fig:Yplot_2}. The new parameter bounds for $\delta_1$, $\delta_2$, and $\delta_3$ are updated as $\mu \pm 3\sigma$, where $\mu$ is the estimated mean and $\sigma$, the estimated standard deviation. 

\begin{figure}
	\centering
	\includegraphics[width=0.4\linewidth,trim={0.0cm 0.0cm 0.0cm 0.0cm},clip]{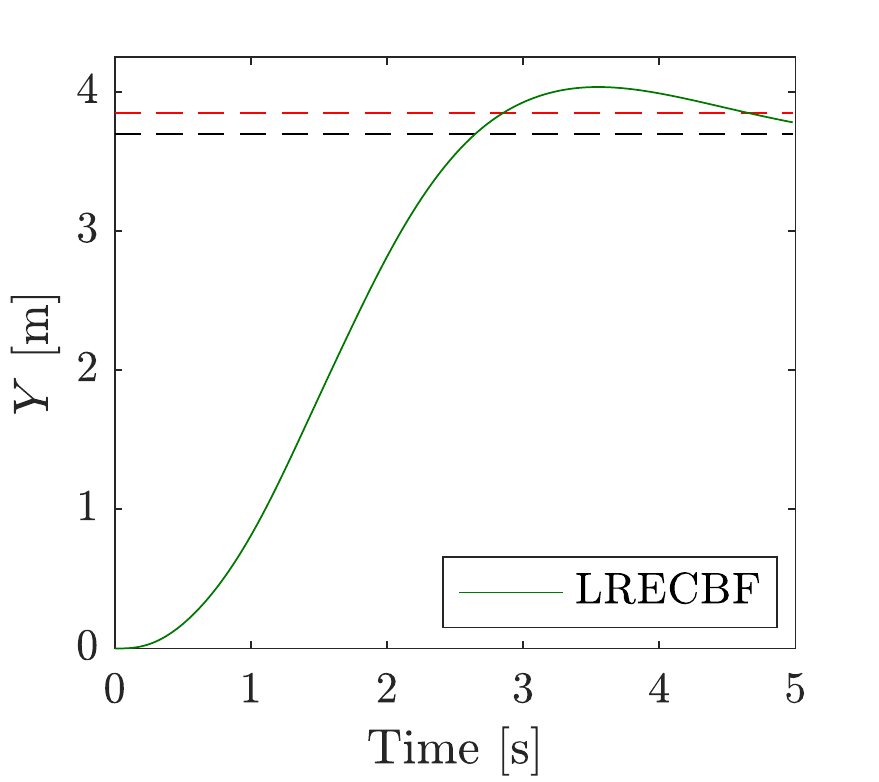}
	\caption{Lane change maneuver with Vehicle \#1 using parameters identified from data obtained with Vehicle \#2.}
	\label{fig:Yplot_wrong}
\end{figure}

Figure~\ref{fig:Yplot_wrong} highlights the risk associated with sudden changes in system dynamics. The figure shows a simulation where the actual system behaves as Vehicle~\#1, but the \textsc{lrecbf} uses parameter estimates from data collected with Vehicle~\#2. In practice, this could happen if the controller learns the vehicle parameters, then the vehicle is unloaded at a stop and driven to another location, and the controller does not reset the parameter estimates. In effect, given that the vehicle must perform a lane change to gather the data required to estimate the parameters, and that this lane change must always be performed safely, the only solution would be to reset the uncertainty bounds to the initial conservative values every time the vehicle stops.

\begin{table}
	\centering
	\caption{Parameter estimation for Example 1.}\label{tbl:pdeltas}
	\begin{tabular}{c | c c c | c c c}
		& \multicolumn{3}{c}{Vehicle \#1} & \multicolumn{3}{|c}{Vehicle \#2} \\
		Param.   & Actual & $\mu$ & $\sigma$       & Actual & $\mu$ & $\sigma$        \\ \hline
		$\dm$    & 0.80   & -     & -              & 1.20   & -     & -               \\
		$\dI$    & 1.15   & 1.202 & 0.037          & 1.05   & 1.048 & 0.022           \\
		$\delta_1$ & 0.70   & 0.706 & 0.009          & 1.35   & 1.357 & 0.005           \\
		$\delta_2$ & 0.60   & 0.595 & 0.003          & 1.35   & 1.351 & 0.006           \\
		$\delta_3$ & 1.35   & 1.386 & 0.073          & 1.35   & 1.354 & 0.062           \\ \hline
	\end{tabular}
\end{table}

\section{Example 2: vehicle longitudinal control} \label{sec:example2}
\begin{figure}
	\centering
	\subfloat[Vehicle model. \label{fig:trucks}]{\includegraphics[width=0.7\linewidth,trim={0.5cm 26.5cm 12cm 1.0cm},clip]{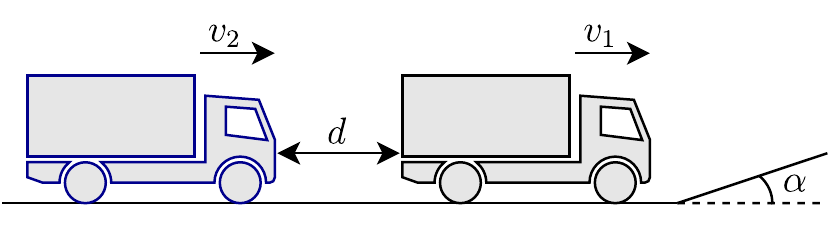}} \\
	\subfloat[Aerodynamic drag reduction.\label{fig:drag}]{\includegraphics[width=0.5\linewidth,trim={0.0cm 0.0cm 0.0cm 0.0cm},clip]{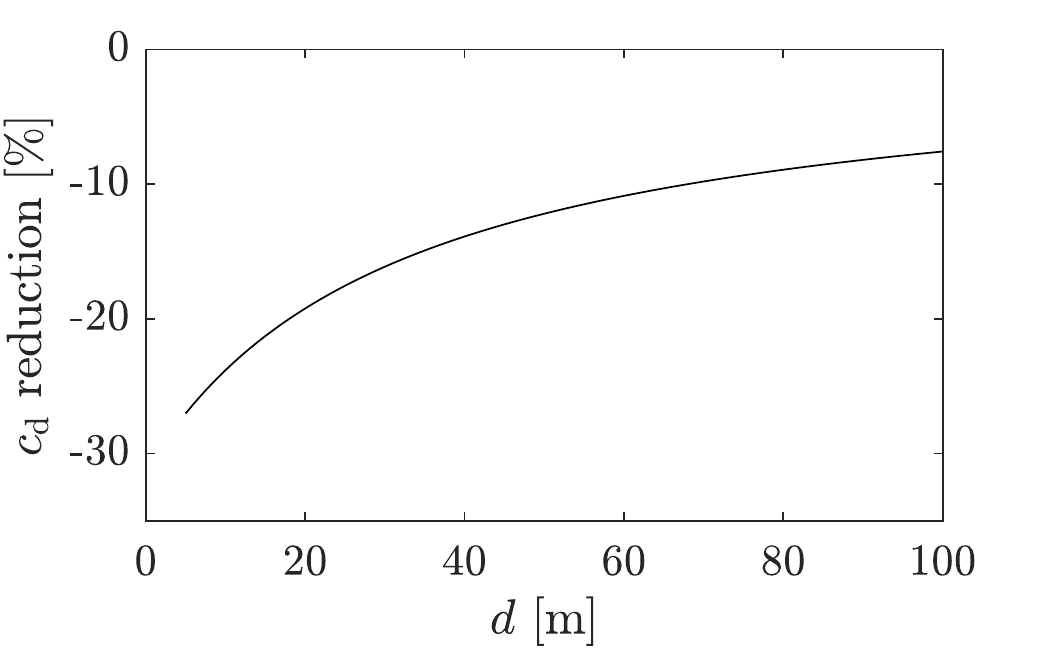}}
	
	\caption{Vehicle platooning model used for Example 2.}
	\label{fig:ACCproblem}
\end{figure}

The second example problem is controlling the trailing vehicle in a two-vehicle platoon, as pictured in Figure~\ref{fig:trucks}. It is assumed that there is no communication between the two vehicles, and that the leading vehicle operates independently of the trailing vehicle. The trailing vehicle can measure its own speed $v_2$, the distance $d$ between the two vehicles, and the road grade $\alpha$. The aerodynamic drag on the trailing vehicle reduces when $d$ reduces~\cite{mcauliffe_track-based_2021}, as shown on Figure~\ref{fig:drag}. Therefore, the control objective is to minimize the distance between the two vehicles, without ever risking a collision.

\subsection{System model \& uncertainty}
The states are $\bx = [v_1, v_2, d]^\top$, the only control action ${\bu = [u]}$ is either a tractive force or a braking force on the second vehicle, and the disturbances are $\bm{\mathrm{v}} = [a_1, \alpha]^\top$. The dynamics for the actual system are
\begin{equation}
	\begin{bmatrix}
		\dot{v_1} \\ \dot{v_2} \\ \dot{d}
	\end{bmatrix} = 
	\begin{bmatrix}
		0 \\ -\frac{1}{2m} \rho v_2^2 \af \cd(d) - g \ct \\ v_1-v_2
	\end{bmatrix} + 
	\begin{bmatrix}
		0 \\ \frac{1}{m} \\ 0
	\end{bmatrix} u + 
	\begin{bmatrix}
		1 & 0 \\ 0 & -g \\ 0 & 0
	\end{bmatrix}
	\begin{bmatrix}
		a_1 \\ \alpha
	\end{bmatrix},
\end{equation}
where
\begin{equation}
	\cd(d) = \cdzero \left( 1- \frac{c_1}{c_2 + d} \right). \label{eq:drag}
\end{equation}
In this model, $m$ is the mass of the trailing vehicle, $\af$ is its frontal area, $\cd$ the aerodynamic drag coefficient, $\ct$ the tire rolling resistance coefficient, $a_1$ is the leading vehicle's acceleration rate, $g$ = 9.81\,m/s\sq, and $\rho$ = 1.225\,kg/m\cub. Full state measurement is assumed for this problem. The actual vehicle parameters are not exactly known, but they are assumed to be centered around nominal values as shown in Table~\ref{tbl:params}. The actual vehicle exhibits a behavior where $\cd(d)$ decreases when $d$ decreases, as per Equation~\eqref{eq:drag} and Figure~\ref{fig:drag}, with the $c_1$ and $c_2$ parameters of Table~\ref{tbl:params}. However, for the system model used in the computation of the nominal and safety controllers, this effect is neglected. Instead, it is assumed that
\begin{equation}
	\cd = \cdzero.
\end{equation}
This assumption is compensated by considering a larger uncertainty on the $\af \cdzero$ parameter of Table~\ref{tbl:params}. Finally, the tractive force is also bound by a maximal power of $\pmax$ = 250\,kW.

\begin{table}
	\centering
	\caption{Uncertain parameters for Example 2.}\label{tbl:params}
	\begin{tabular}{c c c c c}
		Parameter   & Nominal & Actual & Range        & Unit   \\ \hline
		$m$      & 6500    & 5000   & [4500, 8500]  & kg     \\
		$\af \cdzero$ & 4.9     & 4.2    & [3.4, 5.6] & m\sq   \\
		$c_1$     & -       & 10     & -            & m      \\
		$c_2$     & -       & 32     & -            & m      \\
		$\ct$     & 0.006   & 0.007  & [0.005, 0.007]            & -      \\ \hline
		Disturbance  & Nominal & Actual & Range        & Unit  \\  \hline
		$a_1$     & 0       & -    & [-9, 2]       & m/s\sq \\
		$\alpha$    & 0       & -    & [-0.06, 0.06] & rad    \\ \hline
	\end{tabular}
\end{table}

\subsection{Safety controller}
\begin{figure}
	\centering
	\subfloat{\includegraphics[width=0.48\linewidth,trim={0.0cm 0.0cm 0.0cm 0.0cm},clip]{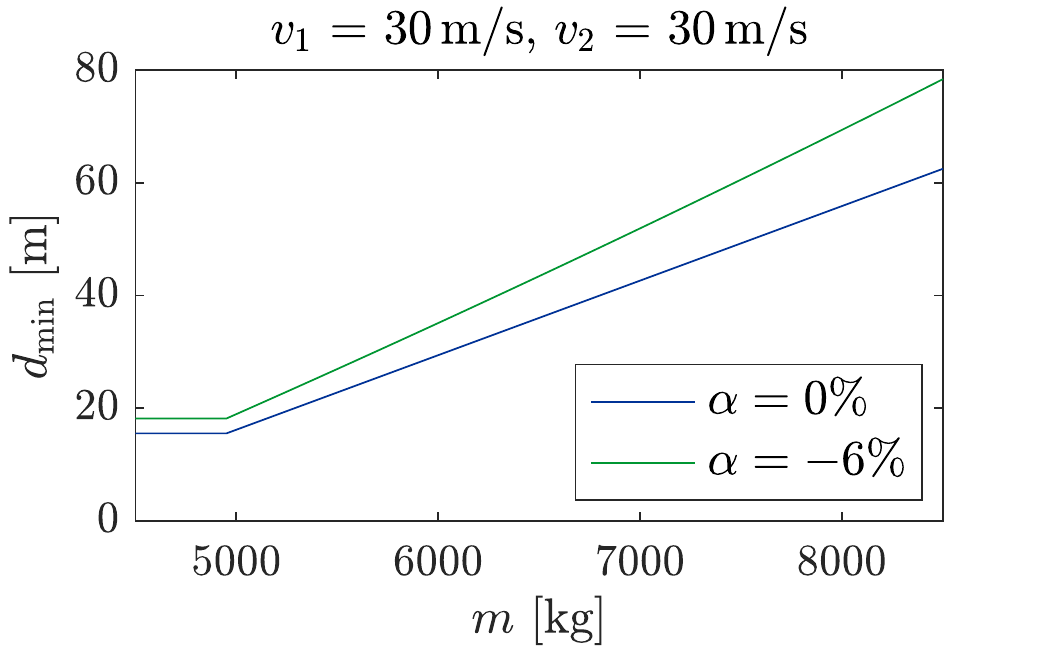}} \quad
	\subfloat{\includegraphics[width=0.48\linewidth,trim={0.0cm 0.0cm 0.0cm 0.0cm},clip]{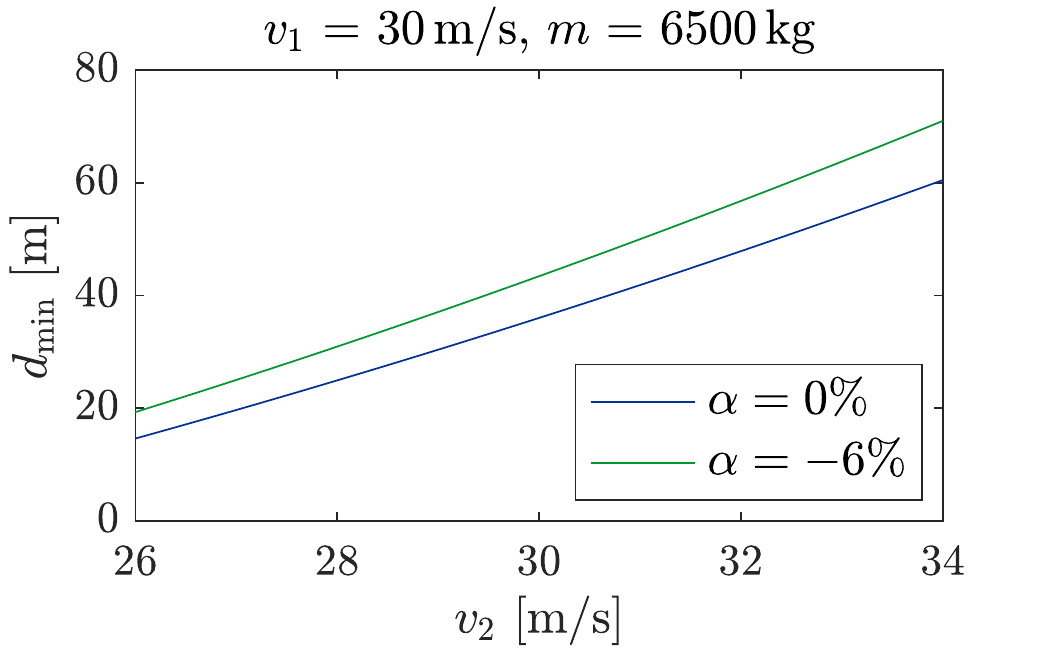}}
	
	\caption{Safe distance $\dmin$ as a function of vehicle mass, road grade, and vehicle speed.}
	\label{fig:dist}
\end{figure}

The safety condition is partially based on the notion of a safe distance $\dmin$. This distance represents the initial distance $d$ such that, if both vehicles come to a full stop with their respective maximum deceleration rate, the final distance $d$ is zero. This distance $\dmin$ is a function of the vehicle speeds, as well as the maximum deceleration rate for the first $\aoe (\alpha)$ and second $\ate (m,\alpha)$ vehicles. It is assumed that the leading vehicle is capable of exploiting a static friction coefficient of $\mus$~=~0.9. Moreover, an additional 0.17\,m/s\sq\ of deceleration due to aerodynamic drag is considered for this vehicle. It is assumed that the trailing vehicle is only capable of exploiting a friction coefficient of $\mus$~=~0.7, and that it is also constrained by a maximum braking force $\Fbmax$. This force is computed assuming that at maximum payload, the vehicle has a maximum deceleration rate of only 4\,m/s\sq, which is just enough to comply with the Federal Motor Vehicle Safety Standards~\cite{us_department_of_transportation_laboratory_2005}. Aerodynamic drag is neglected for the trailing vehicle, which is a conservative assumption. Figure~\ref{fig:dist} shows how both the trailing vehicle mass $m$ and the road slope $\alpha$ influence $\dmin$.
\begin{align}
	\dmin(v_1,v_2,m,\alpha) &= \frac{-v_2^2}{2 \ate (m, \alpha)} + \frac{v_1^2}{2 \aoe(\alpha)}, \\
	\aoe (\alpha) &= -g \alpha - 9, \\
	\ate (m,\alpha) &= -g \alpha - \min \{ \Fbmax m^{-1},0.7g \}.
\end{align} 

Safety is enforced by two control barrier functions, $h_1(\bx)$ and $h_2(\bx)$. The first ensures that the trailing vehicle stays behind $\dmin$, and the second enforces a maximum speed \linebreak $\vmax$ = 32\,m/s.
\begin{align}
	h_1(\bx) &= d-\dmin, \\
	h_2(\bx) &= \vmax - v_2,
\end{align}
For $h_1(\bx)$, a system $\eta(\bx) = [h_1(\bx),\hp_1(\bx)]^\top$ is defined. Then a controller $K_\alpha = [k_1,k_2]$ is obtained from pole placement. This results in the following condition on the control action
\begin{align}
	u &\leq \min_{\bd \in \mathcal{D}} s_1(\bx,\bd), \label{eq:h1}\\
	s_1(\bx,\bd) &:= m\big( k_1(d-\dmin) + k_2(v_1-v_2) \nonumber \\
	&\quad \quad + \tfrac{1}{2m}\rho v_2^2 \af \cdzero + g (\ct + \alpha) + a_1 \big), \\
	\bd &= [m,\af \cdzero, \ct, a_1, \alpha].
\end{align}
For $h_2(\bx)$, an extended class $\Kinf$ function is defined as $-k_3 h_2(\bx)$ where $k_3>0$. Using the condition that $\hp_2(\bx,\bu) \geq -k_3 h_2(\bx),\ \forall \bd \in \mathcal{D}$, a second condition is obtained for the control action 
\begin{align}
	u &\leq \min_{\bd \in \mathcal{D}} s_2(\bx,\bd), \label{eq:h2}\\
	s_2(\bx,\bd) &:= m\big( k_3(\vmax-v_2) + \tfrac{1}{2m}\rho v_2^2 \af \cdzero + g (\ct + \alpha) \big), \\
	\bd &= [m,\af \cdzero, \ct, \alpha].
\end{align}
For both Conditions~\eqref{eq:h1}~and~\eqref{eq:h2}, the minimization problem is much simpler than for Example~1; it can be solved exactly. \\

The base controller $\kn(\bx)$ is a combination of feedforward and feedback signals, where
\begin{align}
	\kn(\bx) &= \uff + \ufb(\bx), \\
	\uff &= -\tfrac{1}{2} \rho \vnom^2 \af \cdzero, \\
	\ufb(\bx) &= \kp(d-\dmin).
\end{align}
The speed $\vnom$ = 30\,m/s is an approximate nominal speed for this problem. $\kp$ is a simple proportional gain. 

\subsection{Parameter identification}
As with the previous example, online parameter identification is performed according to the method presented in Section~\ref{sec:sysID}, where
\begin{align}
	y_{[n]} &= -\tfrac{1}{2 m} \rho ({v_2}_{[n]})^2 \af\cd - g\ct + \tfrac{1}{m}u_{[n]}, \\
	z_{[n]} &= \dot{v_2}_{[n]} + g\alpha_{[n]} + w_{[n]}, \\
	w_{[n]} &\sim \mathcal{N}(0,0.05^2), \\
	\btheta &= [m,\af\cd,\ct]^\top.
\end{align}
Again, noise is injected into the measurements to obtain a more realistic system identification problem.

\subsection{Simulated environment}
For this problem, the road profile is taken from a portion of the \textit{Fleet DNA Long-Haul Representative} drive cycle from the National Renewable Energy Laboratory~\cite{zhang_development_2021}. This defines the road slope to be used during simulation; it has a minimum grade of -6\%. Then, the $v_1(t)$ speed profile is obtained from simulating a vehicle where a PID controller attempts to track a $v_1$ = 30\,m/s velocity. Finally, the simulation for the trailing vehicle can begin, whose details and results are discussed in the next section. 

\subsection{Results \& discussion}
\begin{figure}
	\centering
	\includegraphics[width=1\linewidth,trim={2.0cm 0.5cm 1.5cm 0.5cm},clip]{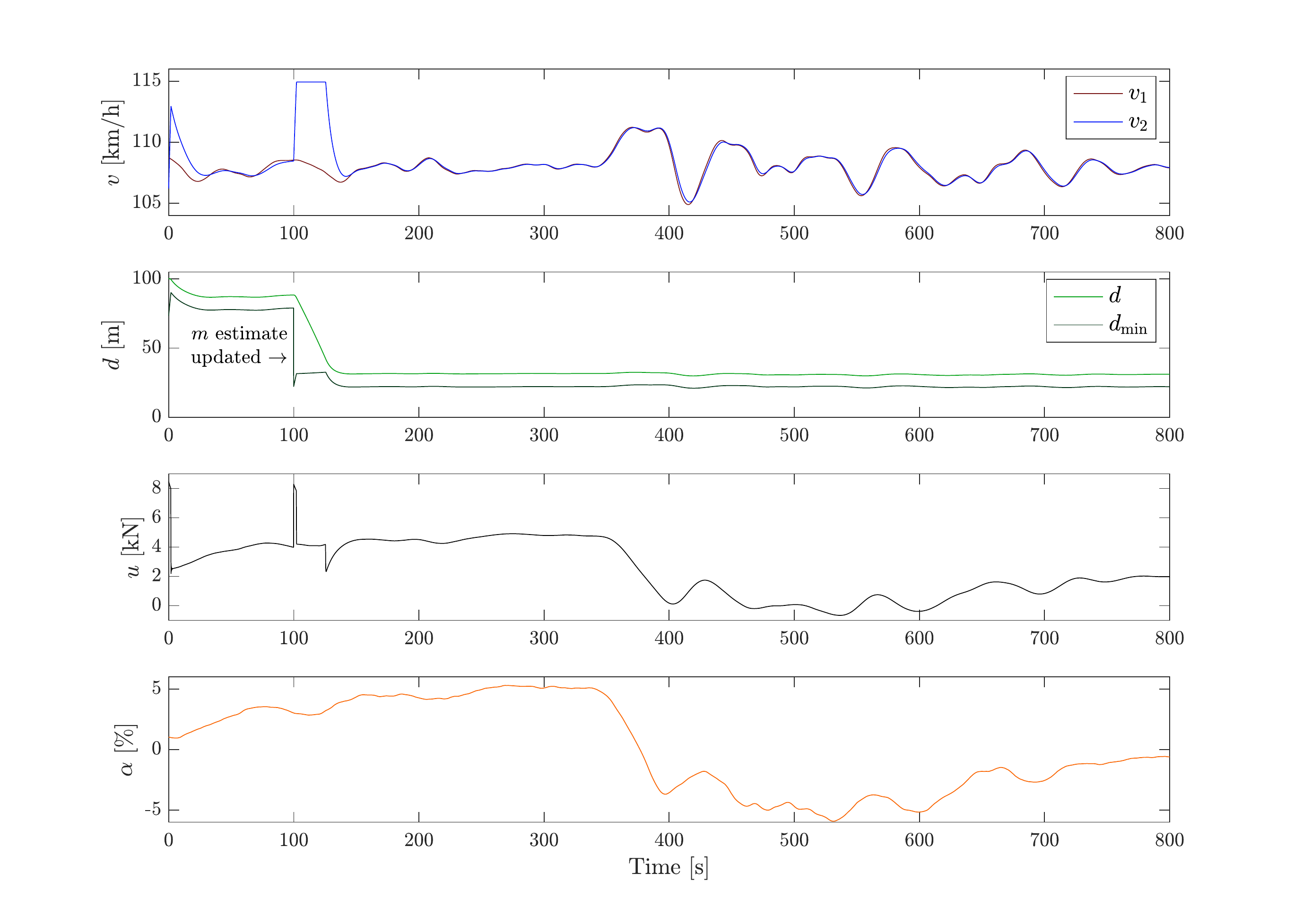}
	\caption{Simulation results for Example 2.}
	\label{fig:simul}
\end{figure}

For this problem, only the robust formulation of the control barrier functions $h_1(\bx)$ and $h_2(\bx)$ are implemented. The simulation begin with $v_2$ = 29.5\,m/s and $d$ = 100\,m. As per Table~\ref{tbl:params}, the initial range for the mass parameter is 4500--8500\,kg. This results in a conservative behavior, where $\dmin\approx$~80\,m for the first section of the simulation, see Figure~\ref{fig:simul}. Using 25\,s of recorded data, the vehicle parameters are estimated online; the results are shown in Table~\ref{tbl:estim}. The resulting variance on the $\af\cd$ and $\ct$ parameters are too high to extract useful information. This is caused by the noise level being too high for the parameters to be identified properly. However, the mass $m$ can be updated to $\mu \pm 3\sigma$, which effectively reduces the range of the parameter. In the simulation, this uncertainty bound is updated at time $t$ = 100\,s, which results in a large increase in system performance. 

\begin{figure}
	\centering
	\includegraphics[width=0.45\linewidth,trim={0.0cm 0.0cm 0.0cm 0.0cm},clip]{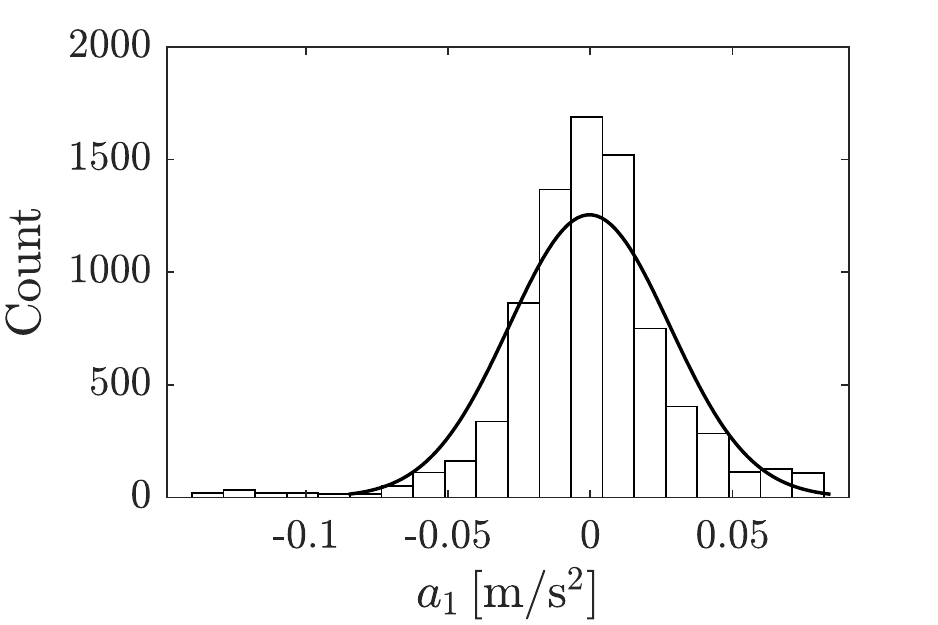}
	\caption{Histogram of the $a_1$ disturbance during the simulation of Figure~\ref{fig:simul}.}
	\label{fig:hist}
\end{figure}

In the simulation, the range for the two disturbances $a_1$ and $\alpha$ are not updated. Likewise, the functions $\aoe(\alpha)$ and $\ate(m,\alpha)$ required to compute $\dmin$ are not updated either. Figure~\ref{fig:hist} shows a histogram of the recorded $a_1$ for the whole duration of the simulation; it also shows a normal distribution fitted on the data. Given that the $\dmin$ and $h_1(\bx)$ safety conditions are defined based on an extreme event---an emergency braking scenario---and that the recorded data does not include such a scenario, it would be impossible to safely update the range of $a_1$ from online learning. Similarly, updating the functions $\aoe(\alpha)$ and $\ate(m,\alpha)$ would require the two vehicles to perform one or several brake tests. Finally, the range for the road slope $\alpha$ is not updated either. This is because even if the vehicle can record the previous values of $\alpha$, this gives no information with regard to the future values. In this problem, it is insufficient to react to the current and previous levels of the disturbances $a_1$ and $\alpha$, as a sudden increase in disturbance could cause a sudden increase in $\dmin$, for which the system would be unable to react to, thereby causing the system to suddenly leave the safe set. For the road slope $\alpha$, one could think of a map-based solution where, given its current position on a predefined trajectory, the system could predict the expected disturbance range for the next 30\,s or so, which leaves enough time to react to.

\begin{table}
	\centering
	\caption{Parameter estimation for Example 2.}\label{tbl:estim}
	\begin{tabular}{c c c c c}
		&                & \multicolumn{2}{c}{Estimation} &      \\
		Parameter & Actual         & $\mu$ & $\sigma$               & Unit \\ \hline
		$m$    & 5000           & 5026  & 57                     & kg   \\
		$\af \cd$ & $\approx$ 3.83 & 3.92  & 0.95                   & m\sq \\
		$\ct$   & 0.007          & 0.006 & 0.011                  & -    \\ \hline
	\end{tabular}
\end{table}

\section{Discussion} \label{sec:discussion}
The first point in Principle~\ref{prop:1} is supported by the results of Example 1. The CBF based on the nominal model is insufficient to keep Vehicle \#1 within state constraints. The results in~\cite{taylor_learning_2020} suggest that a CBF based on a learned nominal model could keep a uncertain system safe, as long as the learned model closely matches the actual system behavior. While this may be the case for some scenarios, this method yields no formal guarantees. We note that in practice, CBFs include a certain amount of conservatism, which depends on the chosen $\alpha$ function. In effect, this function influences the rate at which the system is allowed to evolve toward the set boundary. Therefore, a more conservative CBF may compensate for modeling error. Moreover, as exemplified by Figure~\ref{fig:Yplot_2}, not all model perturbation results in unsafe behavior. Nevertheless, it remains that a nominal model cannot provide safety guarantees, even if it is learned from data. \\

The second point in Principle~\ref{prop:1} is trivially supported by the first point: if a safety condition based on a nominal model can lead the system outside state constraints, then too small an uncertainty bound can certainly do so. 

But perhaps a more interesting discussion concerns learned safety conditions based on non-parametric models. Such models are interesting because the shape of the uncertainty need not be specified in advance, which can lead to a less conservative uncertainty description. Researchers in~\cite{taylor_towards_2021} use data to construct a point-wise uncertainty set that bounds the dynamics, which they use in a robust-CBF setting. Several studies use a Gaussian process (GP)~\cite{rasmussen_gaussian_2006,liu_gaussian_2018} as a non-parametric model to both learn (or correct) a nominal model of the dynamics as well as bound its uncertainty~\cite{fisac_general_2019,hewing_cautious_2020,cheng_end--end_2019, khojasteh_probabilistic_2020,wang_safe_2018}. In effect, GP models provide a probability distribution as state prediction. This distribution can be used to bound the uncertainty on the dynamics~\cite{hewing_simulation_2020}. Despite being a non-parametric model, the distribution obtained from a GP is still dependent on its hyper-parameters, namely the type of kernel function and its chosen parameters. Thus, such hyper-parameters are often optimized through maximum likelihood estimation, see~\cite{fisac_general_2019, hewing_cautious_2020}, much like the parameter estimation presented in this paper. 

Safety conditions based on non-parametric models require special care when first initialized, given that no data is yet collected, thus no safety condition can be computed. One approach is to initially use a nominal model, but restrict the initial safe set to a smaller subset than the entire state constraints~\cite{fisac_general_2019,wang_safe_2018}. In~\cite{fisac_general_2019} for instance, the initial safe set is computed though HJR with assumed conservative bounds on the disturbance. As data gets collected, it is then possible to use the robust safety condition and to iteratively increase the size of the safe set.  Therefore, approaches based on non-parametric models can still meet the condition of Point 2 in Principle~\ref{prop:1}, but additional assumptions are required for the method's initialization. \\

The third point in Principle~\ref{prop:1} is supported by both Examples~1 and 2. In Example 1, the system could become unsafe if the actual dynamics suddenly shift outside the set covered by the estimated uncertainty. In Example 2, the system could become unsafe due to an adversarial disturbance not covered by the estimated uncertainty. These are particularly difficult situations to account for, and they share the same root cause: the data previously collected does not sufficiently describe the future behavior of the system. In Sections~\ref{sec:example1}~and~\ref{sec:example2}, we proposed example solutions to this type of problem. For uncertainty sources that are assumed to remain mostly constant for periods of time, but that are still expected to vary (e.g., the dynamics in Example~1), the controller should predict when a significant change may occur, and reset the corresponding uncertainty bounds to more conservative values. For uncertainty sources that represent rare and extreme events (e.g., the $a_1$ disturbance in Example~2), the controller should never attempt to learn the corresponding uncertainty parameters. 

The change prediction and adaptation mechanism we proposed for Example~1 is simple: the controller resets the uncertainty bounds every time the vehicle stops. Moreover, the controller does not need to continually update the uncertainty parameters as the dynamics are assumed to remain constant when the vehicle operates. For some systems, however, the dynamics could evolve continually, or a sudden change in dynamics could be impossible to predict in advance. These scenarios require a more sophisticated adaptation law if machine learning is to be used to define the safety condition. In~\cite{fisac_general_2019}, researchers propose to constantly evaluate the validity of the learned uncertainty bounds. When the learned model is deemed invalid, a pre-computed safety action is applied, the uncertainty bounds are reevaluated, and finally the updated safe controller comes back online and resumes normal operation. In~\cite{lopez_robust_2021}, the controller constantly identifies the dynamics with set membership identification, and updates the uncertainty parameters accordingly. However, using one of these adaptive methods to guarantee safety in the face of a possibly changing dynamics still requires to make additional assumptions, notably on the magnitude and the rate of these possible changes. Certainly not every scenario is recoverable.

\section{Conclusion}
In this work, we proposed the safe uncertainty-learning principle, which includes three necessary conditions for learning-based control methods to preserve safety guarantees for uncertain dynamical systems. Two example problems were solved where it was shown that not respecting the principle leads to unsafe situations. In summary, it is important to use a robust safety condition---a safety condition solely based on a nominal model is insufficient to obtain formal guarantees, even if this model is learned from data. Moreover, the robust condition must be initialized with sufficiently conservative uncertainty bounds before data can be collected. For safety conditions based on non-parametric models, this requires special attention. For instance, an adequate method is to reduce the size of the initial safety set, and allow the controller progressively more freedom as the dynamics and disturbances are identified. Also, it was shown that not all uncertainty bounds should be learned. For some problems, the safety condition is defined by an extreme and rare event whose magnitude cannot be estimated from previously collected data. Fixed uncertainty bounds must be employed to account for such scenarios. For systems where the dynamics are expected to change, the parameter update law must be carefully devised. In particular, the capacity of the system to respond to the expected change is a key factor. In conclusion, the safe uncertainty-learning principle offers an approach to evaluate whether a learning method employed during the control of an uncertain dynamical system preserves safety guarantees. 

\bibliographystyle{IEEEtranMdoi}
\bibliography{main}

\begin{IEEEbiography}[{\includegraphics[width=1in,height=1.25in,clip,keepaspectratio]{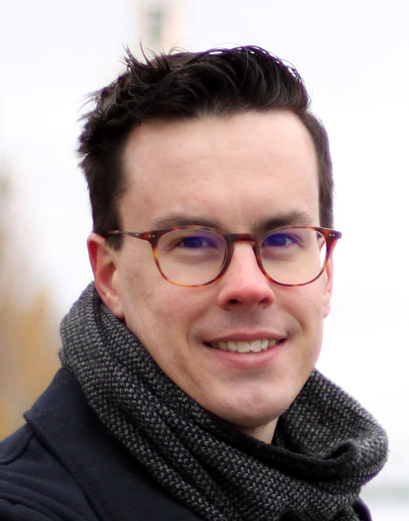}}]{Marc-Antoine Beaudoin}
	is a doctoral student at the Intelligent Automation Laboratory of the Centre for Intelligent Machines at McGill University. He obtained a Bachelor of Engineering degree from Université de Sherbrooke in 2015, and a Master of Applied Science degree from the University of Toronto in 2018, both in Mechanical Engineering. He is currently pursuing the Ph.D. degree in Electrical and Computer Engineering at McGill University. Prior to his doctoral studies, Mr. Beaudoin was a practicing engineer in the electric vehicle industry. His research areas include the machine learning control of dynamical systems, with a focus on electric and autonomous vehicles.
\end{IEEEbiography}
\begin{IEEEbiography}[{\includegraphics[width=1in,height=1.25in,clip,keepaspectratio]{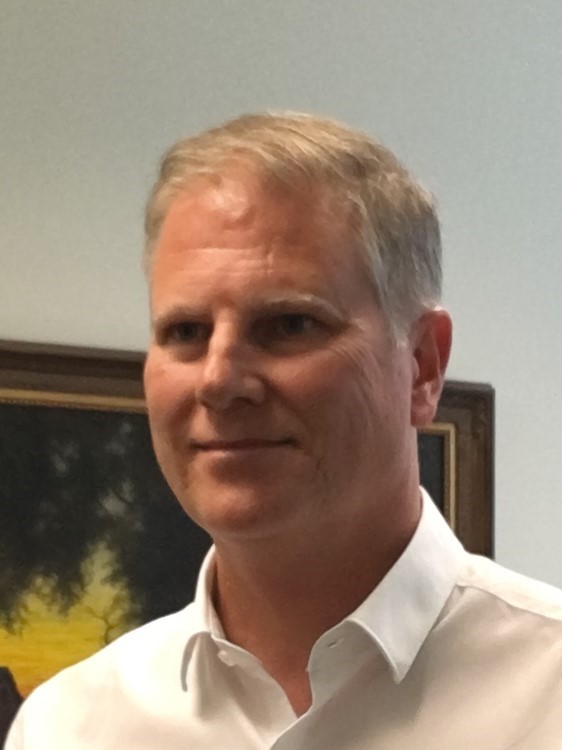}}]{Benoit Boulet}
	(Senior Member, IEEE) is Professor in the Department of Electrical and Computer Engineering at McGill University which he joined in 1998, and Director of the McGill Engine, a Technological Innovation and Entrepreneurship Centre. He is Associate Vice-Principal of McGill Innovation and Partnerships and was Associate Dean (Research \& Innovation) of the Faculty of Engineering from 2014 to 2020. Professor Boulet obtained a Bachelor's degree in applied sciences from Université Laval in 1990, a Master of Engineering degree from McGill University in 1992, and a Ph.D. degree from the University of Toronto in 1996, all in electrical engineering. He is a former Director and current member of the McGill Centre for Intelligent Machines where he heads the Intelligent Automation Laboratory. His research areas include the design and data-driven control of electric vehicles and renewable energy systems, machine learning applied to biomedical systems, and robust industrial control.
\end{IEEEbiography}

\end{document}